\documentclass[pdflatex,sn-mathphys-num]{sn-jnl}

\usepackage{colortbl}
\usepackage{graphicx}%
\usepackage{multirow}%
\usepackage{amsmath,amssymb,amsfonts}%
\usepackage{amsthm}%
\usepackage{mathrsfs}%
\usepackage[title]{appendix}%
\usepackage{xcolor}%
\usepackage{textcomp}%
\usepackage{manyfoot}%
\usepackage{booktabs}%
\usepackage{algorithm}%
\usepackage{algorithmicx}%
\usepackage{algpseudocode}%
\usepackage{listings}%
\usepackage{utfsym}%

\theoremstyle{thmstyleone}%
\theoremstyle{thmstyletwo}%

\theoremstyle{thmstylethree}%

\begin{document}


\title[HECLIP]{HECLIP: Histology-Enhanced Contrastive Learning for Imputation of Transcriptomics Profiles}

\author[1]{\fnm{Qing} \sur{Wang}}

\author[2]{\fnm{Wen-jie} \sur{Chen}}

\author[3]{\fnm{Bo} \sur{Li}}

\author[4]{\fnm{Jing} \sur{Su}}

\author[5,6]{\fnm{Guangyu} \sur{Wang}}

\author*[1]{\fnm{Qianqian} \sur{Song}}\email{qsong1@ufl.edu}

\affil[1]{\orgdiv{Department of Health Outcomes and Biomedical Informatics}, \orgname{College of Medicine, University of Florida}, \orgaddress{\postcode{32611}, \state{FL}, \country{USA}}}

\affil[2]{\orgdiv{School of Biological and Behavioural Sciences}, \orgname{Queen Mary University of London}, \orgaddress{\street{} \postcode{E1 4NS}, \state{London}, \country{UK}}}

\affil[3]{\orgdiv{Department of Computer and Information Science}, \orgname{University of Macau}, \orgaddress{\street{Taipa, Macau SAR}, \postcode{999078}, \country{China}}}

\affil[4]{\orgdiv{Department of Biostatistics and Health Data Science}, \orgname{Indiana University School of Medicine}, \orgaddress{\postcode{46202}, \state{IN}, \country{USA}}}

\affil[5]{\orgdiv{Center for Bioinformatics and Computational Biology}, \orgname{Houston Methodist Research Institute}, \orgaddress{\postcode{77030}, \state{TX}, \country{USA}}}
\affil[6]{\orgdiv{Department of Cardiothoracic Surgery}, \orgname{Weill Cornell Medicine, Cornell University}, \orgaddress{\state{NY}, \country{USA}}}


\abstract{Histopathology, particularly hematoxylin and eosin (H\&E) staining, plays a critical role in diagnosing and characterizing pathological conditions by highlighting tissue morphology. However, H\&E-stained images inherently lack molecular information, requiring costly and resource-intensive methods like spatial transcriptomics to map gene expression with spatial resolution. To address these challenges, we introduce HECLIP (Histology-Enhanced Contrastive Learning for Imputation of Profiles), an innovative deep learning framework that bridges the gap between histological imaging and molecular profiling. HECLIP is specifically designed to infer gene expression profiles directly from H\&E-stained images, eliminating the need for expensive spatial transcriptomics assays. HECLIP leverages an advanced image-centric contrastive loss function to optimize image representation learning, ensuring that critical morphological patterns in histology images are effectively captured and translated into accurate gene expression profiles. This design enhances the predictive power of the image modality while minimizing reliance on gene expression data. Through extensive benchmarking on publicly available datasets, HECLIP demonstrates superior performance compared to existing approaches, delivering robust and biologically meaningful predictions. Detailed ablation studies further underscore its effectiveness in extracting molecular insights from histology images. Additionally, HECLIP’s scalable and cost-efficient approach positions it as a transformative tool for both research and clinical applications, driving advancements in precision medicine. The source code for HECLIP is openly available at https://github.com/QSong-github/HECLIP.}


\maketitle

\section{Introduction}
Histopathology is widely recognized as a standard for identifying and characterizing diverse pathological conditions. Central to histopathological procedures is tissue staining, which differentiates intracellular components to facilitate visual interpretation. Among the staining techniques, hematoxylin and eosin (H\&E) staining is the most widely used. It exploits the contrasting affinities of acidic eosin and basic hematoxylin dyes to highlight tissue morphology \cite{feldman2014tissue}, providing pathologists with essential visual cues for diagnostic decision-making \cite{li2024gene}. Despite its widespread use and diagnostic value, H\&E-stained images inherently carry limited molecular information \cite{bonasia2015intra}, necessitating the expertise of skilled pathologists to interpret nuanced features.

The spatial organization of gene expression within tissues plays a pivotal role in understanding complex biological processes and disease mechanisms. Spatial transcriptomics (ST) has emerged as a revolutionary technology that integrates spatial resolution with gene expression profiling, offering unparalleled insights into tissue heterogeneity and the microenvironment. However, the practical application of ST remains constrained by its high cost and specialized equipment requirements, which limit accessibility for routine clinical or research use \cite{wang2023spatial, sharma2024quantitative}. In contrast, histological imaging is an established and cost-effective technique capable of capturing tissue structure and morphology with high resolution. Computational approaches  \cite{kolodziejczyk2015technology, dorn2016assessment} that infer spatial gene expression from histology images present a promising alternative to ST. These approaches bridge the gap between histology and transcriptomics by leveraging advanced deep learning models, providing a scalable and efficient solution that integrates molecular and morphological information, thereby advancing precision medicine \cite{bai2024spatially,moses2022museum,rao2021exploring}.

Recent studies have developed different methods to predict gene expression from histology images, demonstrating the potential of these computational approaches in reducing the reliance on expensive sequencing technologies for spatially resolved gene expression profiling. For instance, CLIP \cite{radford2021learning} leverages contrastive learning to align image and text modalities, enabling applications such as cross-modal retrieval and classification. ST-Net \cite{he2020integrating} uses deep learning to predict local gene expression directly from H\&E-stained images, while BLEEP \cite{xie2024spatially} employs a bi-modal embedding framework to map paired image and gene expression data into a unified embedding space. HisToGene \cite{pang2021leveraging}, another advanced model, utilizes a Vision Transformer to capture spatial dependencies in histological data, improving gene expression predictions by integrating structural context.

Despite significant advancements, current methods still face limitations in accuracy and reliability. The accurate prediction of gene expression patterns from histological images is challenged by the biological complexity of tissues, where factors such as cell type and microenvironmental influences play critical roles. Overcoming these challenges requires further innovation to address both technical and biological hurdles, therefore achieving robust and biologically meaningful predictions. Such advancements will pave the way for the broader adoption of these technologies in both clinical and research settings. 

In this paper, we present the HECLIP model, which leverages an innovative image-centric contrastive loss to optimize multimodal representation learning. By designing a tailored image-centric loss function, HECLIP enhances the representation capabilities of histological images, enabling more accurate predictions of transcriptomic data. This customized loss function is versatile and adaptable, making it applicable to a wide range of multimodal contrastive learning tasks. Extensive evaluations across multiple datasets demonstrate that HECLIP not only achieves robust performance but also consistently outperforms existing models in different datasets and scenarios, highlighting its effectiveness and broad applicability.

\section{Method}\label{sec3}
\begin{figure*}[!htb]
\centering
\includegraphics[scale = 0.458]{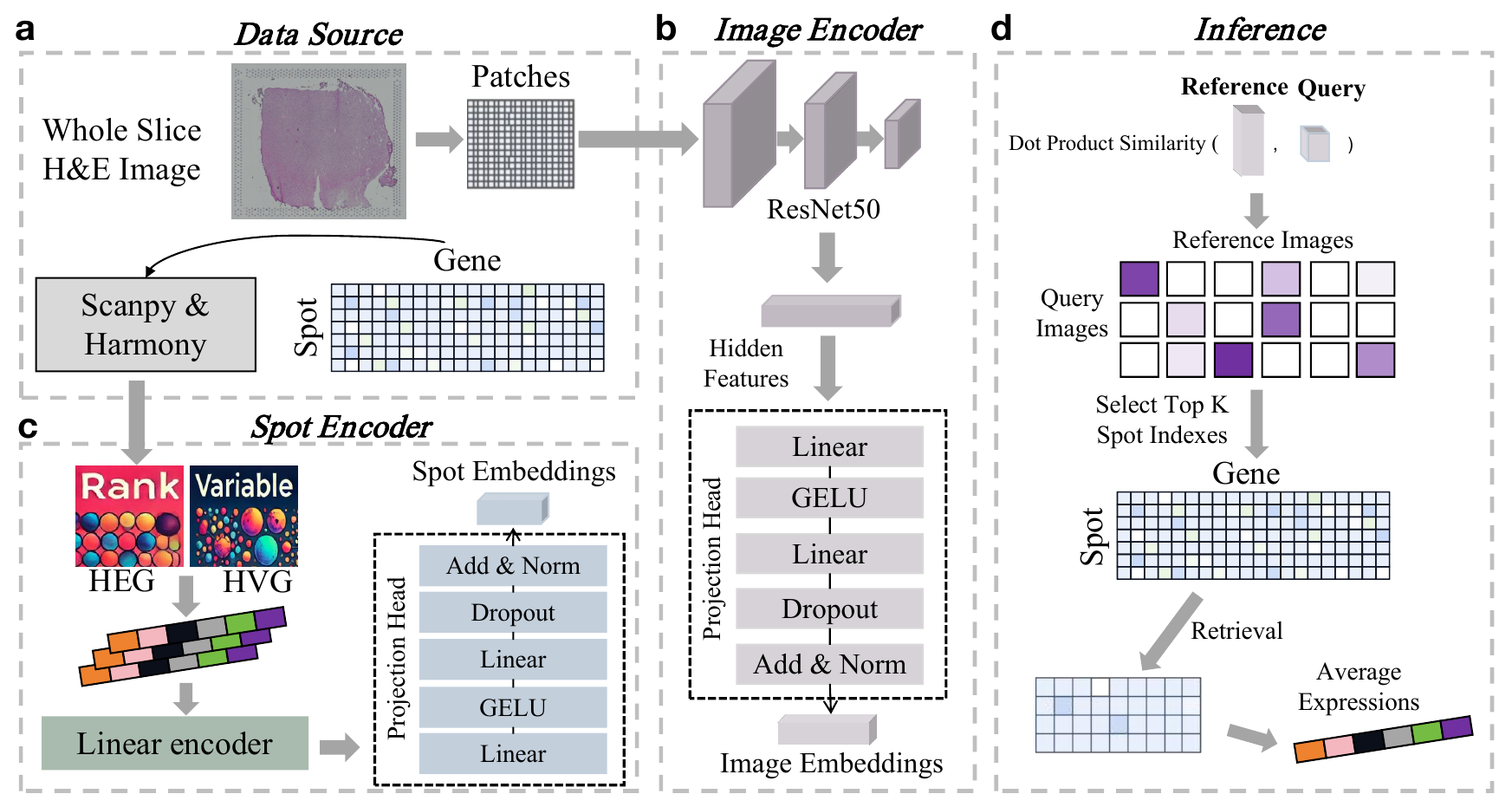}
\caption{ Overview of the HECLIP framework for transcriptomics imputation from histological images.}
\label{model}
\end{figure*}

We present HECLIP (Histology-Enhanced Contrastive Learning for Imputation of Profiles), a deep learning framework (Fig. \ref{model}) designed to infer spatial gene expression profiles directly from hematoxylin and eosin (H\&E)-stained histological images. Fig. \ref{model}a illustrates the data preprocessing steps, where whole-slide histological images are divided into patches 256 × 256 pixels), and corresponding transcriptomic data is normalized and prepared using Scanpy. This preprocessing aligns histological image features with transcriptomic profiles \cite{yuan2023sodb, gao2021delineating}, enabling effective integration. For the image patches, the Image Encoder module (Fig. \ref{model}b) employs a ResNet-50 backbone combined with linear layers, GELU activations, normalization, and dropout layers to extract high-dimensional image embeddings. In parallel, transcriptomic data focusing on highly expressed genes (HEG) and highly variable genes (HVG) is processed by the Spot Encoder module (Fig. \ref{model}c) to generate transcriptomic embeddings at spatially resolved spots. These embeddings are aligned with the image features within a shared embedding space through contrastive learning, facilitating the seamless integration of molecular and morphological data. During the inference stage (Fig. \ref{model}d), embeddings from query image patches (e.g., new, unseen images) are compared to reference embeddings (e.g., training images) to identify the most relevant reference spots. The top-K similar spots are selected, and their gene expression profiles are retrieved and averaged to impute the spatial transcriptomic profiles for the query patches. In this way, HECLIP provides robust and biologically meaningful predictions, bridging the gap between histology and transcriptomics while offering a scalable and cost-effective alternative to traditional spatial transcriptomics methods. 

\begin{table*}[]
\setlength{\tabcolsep}{1mm}{
\begin{tabular}{c|ccc} \hline \hline
Datasets & \begin{tabular}[c]{@{}c@{}}Training size \\ (reference)\end{tabular} & \begin{tabular}[c]{@{}c@{}}Testing size \\ (query)\end{tabular} & Gene size \\ \hline
GSE240429\_HVG      & 6992  & 2277 & 3467 \\
GSE240429\_HEG      & 6992  & 2277 & 3511 \\
GSE245620\_HVG      & 14975 & 4992 & 3508 \\
GSE245620\_HEG      & 14976 & 4992 & 3403 \\
spatialLIBD\_1\_HVG & 12679 & 4789 & 3376 \\
spatialLIBD\_1\_HEG & 12679 & 4789 & 3468 \\
spatialLIBD\_2\_HVG & 25538 & 4110 & 3405 \\
spatialLIBD\_2\_HEG & 25538 & 4110 & 3615 \\ \hline \hline
\end{tabular}}
\caption{Details of datasets. Gene size refers to the number of selected genes from their corresponding gene expression data.}
\label{dtdt}
\end{table*}

\subsection{Datasets and Preprocessing}
To validate the effectiveness of HECLIP, we have evaluated its performance using the GSE240429 dataset \cite{andrews2024single} (available at https://www.ncbi.nlm. nih.gov/geo/query/acc.cgi?acc=GSE 240429), the GSE245620 dataset \cite{andrews2024single} (available at https://www. ncbi.nlm.nih.gov/geo /query/acc.cgi?acc=GSE245620), and the spatialLIBD dataset \cite{maynard2021transcriptome} (split into two datasets and available at https://research.libd .org/spatialLIBD/). These datasets, summarized in Table \ref{dtdt}, are profiled using the 10x Genomics Visium platform.

The GSE240429 and GSE245620 datasets focus on the immunological status of healthy human livers and those affected by primary sclerosing cholangitis (PSC), a chronic liver disease characterized by bile duct inflammation and damage. Both  datasets consist of four consecutive thick sections of human liver tissue from neurodegenerative donor livers deemed suitable for transplantation.

The spatialLIBD dataset is profiled with human dorsolateral prefrontal cortex (DLPFC) slices, which includes regions spanning six neuronal layers plus white matter, covering three subjects with two pairs of spatially adjacent replicates per subject. This dataset comprises 12 slices in total, which were split into two subsets for our analysis: spatialLIBD 1 (4 slices) and spatialLIBD 2 (8 slices), based on sample continuity. Each slice covers all six cortical layers and white matter to ensure comprehensive spatial representation.

For each spot, we extract image patches from the whole-slide H\&E stained histology images and paired them with corresponding gene expression profiles. Gene expression data is normalized and log-transformed using Scanpy \cite{wolf2018scanpy}. Spot coordinates $(x, y)$ are used to form patch-spot pairs, where the patch vertices are calculated as $(x-128, y-128), (x-128, y+128), (x+128, y+128), (x+128, y-128)$, and $(x+128, y-128)$, in pixels.

To evaluate HECLIP’s prediction capability across slices, we use one slice in every dataset as the test set (query) while the remaining slices are used for training (reference). For each spot, we test both highly expressed gene (HEG) and highly variable genes (HVG). For HEG, we rank genes in each spot based on expression levels, selecting the top 3,500 genes across all spots. For HVG, we identify the most variable genes per slice, combine them across slices, and select 3,500 genes for training and prediction. Batch effects between slices are corrected using the Harmonypy \cite{korsunsky2019fast} package to account for technical variability. This experimental setup enables a robust evaluation of HECLIP’s performance across diverse datasets.

\subsection{Model input}
Let $D = \{ (p_1,s_1),(p_2,s_2), ... , (p_n,s_n) \}$ represents the training dataset with $n$ patch-spot pairs, where $p_i\in R^{256\times256}$ is the input of image patch and $s_i\in R^d$ denotes the gene expression sequence. $d$ refers to the number of HEG or HVG (gene size). We also use paired data augmentation approach to improve data diversity. Specifically, we perform random flipping and rotations on each patch, which is commonly used in the visual field. However, we perform two random operations on a patch in one epoch, while the spot remains unchanged. Therefore, we get double the amount of data in each epoch. Ablation experiments are performed to verify the effectiveness of this strategy.

\subsection{Embedding modules}
The image embedding module is the classic ResNet50 \cite{he2016deep,xie2017aggregated} model with a projection head. We use the pre-trained ResNet50 weights from the 'timm' package, while the projection head weights are randomly initialized. A batch of image patches is denoted as $P \in \mathbb{R}^{|B| \times 256 \times 256 \times 3}$, where $|B|$ represents the image batch size, 256 is the patch size, and 3 is the RGB channel. Specifically, ResNet50 consists of four residual block modules (Block1-Block4), each containing multiple convolution layers that further reduce the spatial size and increase the number of channels. Each residual block extracts more robust features through residual connections and convolution operations. First, through the processing of convolution layers and pooling layers, the dimensions (height and width) of images are gradually reduced while the depth (number of channels) of image features are increased. Then the image features pass through the convolution blocks inside ResNet50. After the last convolution block, the feature map size obtained is $(|B|, 8, 8, 2048)$. The global average pooling layer performs average pooling on the $8x8$ feature map on each channel and outputs an embedding of size $(|B|)$. After the pooling operation, the final image representation $z_p \in \mathbb{R}^{|B| \times 2048}$ (the shape is flattened) is obtained.

Projection Head consists of linear projection, nonlinear activation, layer normalization (LayerNorm), residual connection and dropout. First, there is a fully connected linear layer with the GELU activated function that maps the $h_{p}$ to embedding space. Next is a random dropout layer (dropout rate: 0.1) to relieve overfitting. Finally, there is a normalization layer to normalize the output and stabilize the training process. The formulas are as follows:

\begin{equation}
h_1 = GELU( W_1 z_p + b_1 )
\label{f1}
\end{equation}

\begin{equation}
h_2 = Dropout(W_2 h_1 + b_2 )
\label{f2}
\end{equation}

\begin{equation}
h_{p} = LayerNorm(h_1 + h_2) 
\label{f3}
\end{equation}

where $h_{p} \in \mathbb{R}^{|B|\times d_o}$ is the image embedding and $d_o$ is the output dimension. For the embedding extraction module of spot data, considering its relatively low data complexity, we adopt a shallow linear network to embed the spots into the $d_o$ dimension, followed by a projection head similar to the image embedding module. Spot embedding can be expressed as $h_{s} \in \mathbb{R}^{|B|\times d_o}$ with $d_o$ dimension of feature.

\begin{algorithm}
\label{sf}
\caption{}
\textbf{Input:} \\ patch embedding ($h_{p} \in \mathbb{R}^{|B|\times d_o}$) \\  spot embedding ($h_{s} \in \mathbb{R}^{|B|\times d_o}$)
\\
\textbf{Output:} loss value ($l \in \mathbb{R}$)
\begin{algorithmic}[1]
    \Function{cross\_entropy}{$logits$, $targets$}
        \State $\text{logits} \gets \text{LogSoftmax(logits, dim=-1)}$ 
        \State $\text{\small{CE\_loss}} \gets (-\text{targets} * \text{logits}).\text{sum(1)}$
        \State \Return $\text{CE\_loss}$
    \EndFunction
    
    \Function{image\_centric\_clip\_loss}{$h_{p}$, $h_{s}$}
        \State $\text{logits} \gets (h_{s} @ h_{p}.T) / \text{temperature}$ 
        \State $\text{sim\_img} \gets h_{p} @ h_{p}.T$
        \State $\text{targets} \gets \text{sim\_img} / \text{temperature}$  
        \State $\text{targets} \gets \text{Softmax(targets)}$
        \State $\text{\small{CE\_loss}} \gets \text{\small{CROSS\_ENTROPY}}(\text{\small{logits.T}}, \text{\small{targets.T}})$
        \State $l \gets \text{CE\_loss.mean()}$
        \State \Return $l$
    \EndFunction
\end{algorithmic}
\end{algorithm}

\subsection{Loss function}
In this work, we have used an innovative contrastive learning loss in HECLIP, specifically tailored to optimize the model parameters for our unidirectional task. Traditional contrastive loss functions, as used in the CLIP model, balance the two modalities (e.g., images and text) equally during loss calculation, making them well suited for bidirectional tasks such as image-to-text or text-to-image mapping. However, our application focuses exclusively on predicting gene expression from histological images, a unidirectional objective. To address this, we prioritize the optimization of the image encoder in HECLIP, enhancing its ability to generate highly informative embeddings from histological images while aligning accurately with the corresponding spot profiles. We achieve this through a simple yet effective strategy: reducing or completely removing the impact of spot-based loss during training. Specifically, we implement an algorithm (detailed in Algorithm 1) that excludes spot-based loss from the training process. During training, 80\% of the data is used for model training, while the remaining 20\% is reserved for testing model performance. The model parameters associated with the lowest test loss are saved as the final configuration of the model. This approach allows the model to focus on refining the image encoder, ensuring accurate prediction of gene expression profiles solely from histological images.

\subsection{Inference stage}
During the inference phase, our approach diverges from the traditional CLIP \cite{radford2021learning} model, which retrieves similar samples from embeddings of opposite modalities (e.g., using image embeddings to match text embeddings). Instead, we focus on the optimized image encoder to exclusively utilize image modality embeddings for retrieval. 

This process begins by extracting image embeddings from both the training and test data using the image encoder with fixed parameters, forming the reference set. Next, the test set images are input into the image encoder to generate the query set embeddings. For each query patch, we calculate the dot product similarity between its embedding with each embedding in the reference set, ranking the results by similarity scores.

A predefined value of \( K \) is then used to select the \( K \) most similar reference patches for each query patch. The corresponding labels of the selected reference patches are retrieved, and the predicted gene expressions for the query patches are determined by averaging the gene expression profiles of these \( K \) reference patches. This approach leverages the optimized image encoder to ensure accurate and robust predictions based on the similarity of image embeddings.

\subsection{Evaluation Metrics}
We have used several popular evaluation metrics in the experiments, including Root Mean Square Error (RMSE) and Structural Similarity Index (SSIM) \cite{li2022benchmarking}.

$RMSE$ measures the deviation between the predicted gene expressions and the actual gene expressions within each spot. The smaller the RMSE, the better the prediction performance of the model.

\begin{equation}
R M S E=\sqrt{\frac{1}{M} \sum_{j=1}^M\left(\tilde{e}_{i j}-e_{i j}\right)^2}
\end{equation}

where $e_{i j}$ and $\tilde{e}_{i j}$ are the normalized spatial expression of gene $i$ in spot $j$ in the ground truth and the predicted result, respectively. $SSIM$ measures the similarity between predicted and true gene expressions across spots. The value of SSIM ranges between -1 and 1. SSIM values closer to 1 indicates more accurate predictions. Following the procedures of \cite{li2022benchmarking}.
We scaled the expression matrix as follows:

\begin{equation}
e_{i j}^{\prime}=\frac{e_{i j}}{\max \left(\left\{e_{i 1}, \ldots, e_{i M}\right\}\right)}
\end{equation}

where $e_{i j}$ denotes the expression of gene $i$ in spot $j$, and $M$ is the total number of spots. Then we calculate the SSIM value as follows:

\begin{equation}
S S I M=\frac{\left(2 \tilde{u}_i u_i+C_1^2\right)\left(2 \operatorname{cov}\left(e_i^{\prime}, \tilde{e}_i^{\prime}\right)+C_2^2\right)}{\left(\tilde{u}_i^2+u_i^2+C_1^2\right)\left(\tilde{\sigma}_i^2+\sigma_i^2+C_2^2\right)}
\end{equation}

where $\mu_i$ and $\tilde{u}_i$ are the average expression value of gene $i$ in the ground truth and the predicted result, respectively; and $\sigma_i$ and $\tilde{\sigma}_i$ are the s.d. of the ground truth and the predicted result, respectively. $\operatorname{cov}(\cdot)$ is the covariance. The $C_1$ and $C_2$ are small constants to stabilize the calculation. We also used the top gene hit rate $Hit@T$ (inspired by extreme multi-label classification \cite{wang2023gudn}).

\begin{equation}
Hit @ T=\frac{1}{N} \sum_{i=1}^N I\left(\text { pred }_i \cap \text { true }_i \neq \emptyset\right),
\end{equation}

where $I$ is the indicator function, which equals 1 if the intersection of the two sets is non-empty (i.e., if there is at least one common index) and 0 otherwise. $T$ specifies the  selection range, for example, if  $T$ = 3 , it means selecting the top 3 highest-expressed genes and checking the overlap between the predicted top-expressed genes and the actual top-expressed genes.

\section{Results}

\subsection{Benchmarking experiments demonstrate superior performance of HECLIP}
\begin{figure*}[!htb]
\centering
\includegraphics[scale = 0.518]{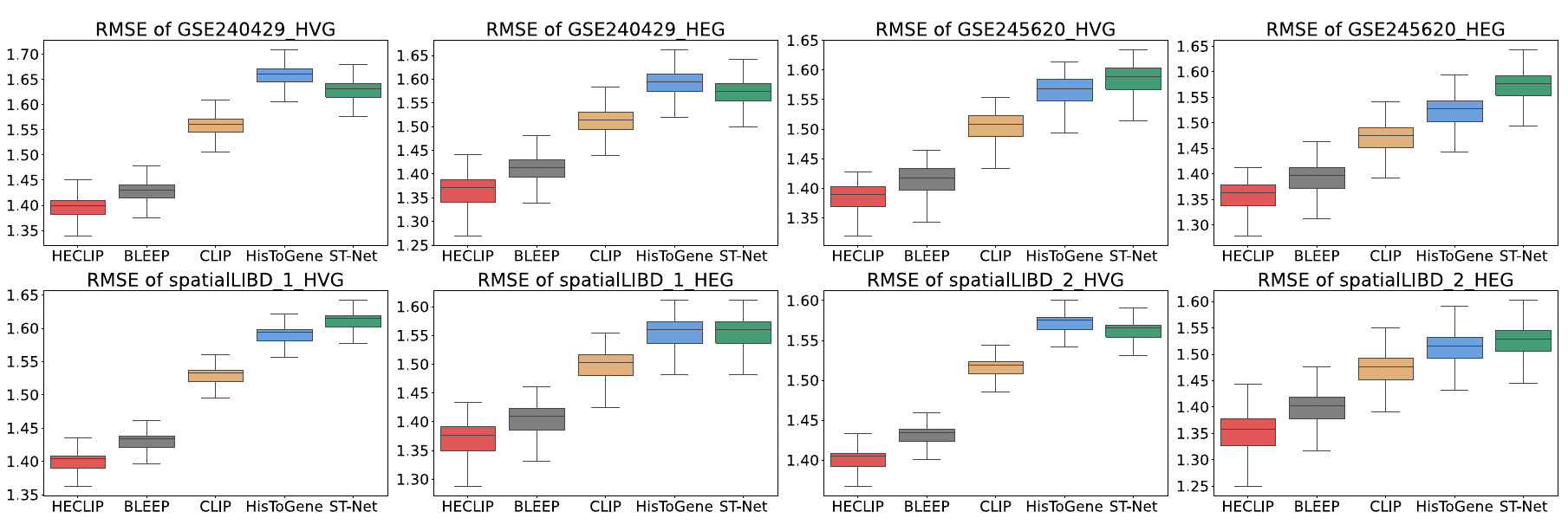}
\caption{Comparison of methods for predicting transcriptomics from histology images based on RMSE metrics.}
\label{rmse}
\end{figure*}

To evaluate the performance of HECLIP compared to existing methods, we conduct benchmarking experiments on multiple publicly available datasets (see Datasets and Preprocessing). The benchmarking results demonstrate that HECLIP consistently outperforms other methods across all datasets in both HVG and HEG scenarios, as evidenced by the SSIM (Supplementary Fig. 1) and RMSE (Fig. \ref{rmse}) metrics.
Specifically, in the GSE240429\_HVG and GSE245620\_HVG datasets, the median RMSE values for HECLIP were 1.40 and 1.39, respectively, with corresponding mean RMSE of 1.39 and 1.37, outperforming other models. For the SSIM metric, HECLIP also excels, achieving a median SSIM of 0.007 and of 0.011 in the GSE240429\_HVG and GSE245620\_HVG dataset, higher than BLEEP and CLIP. This trend is also observed in the spatialLIBD datasets. For example, in spatialLIBD\_2\_HEG, HECLIP achieved a median SSIM of 0.0285 and a mean SSIM of 0.048. Moreover, HECLIP exhibited lower variability in performance, particularly in datasets such as GSE240429\_HEG and spatialLIBD\_2\_HEG, as shown in the box plots for RMSE and SSIM. This indicates that HECLIP is not only more accurate but also more stable and reliable. In contrast, other models like BLEEP and CLIP demonstrated lower SSIM values, particularly in datasets such as spatialLIBD\_2, while HisToGene and ST-Net showed consistently poor predictive performance overall.

The results for Hit@T are presented in Table \ref{hk}. HECLIP consistently achieved the highest accuracy across all Hit@T values, clearly demonstrating its superior predictive capability. For instance, on the spatialLIBD\_1\_HVG dataset, HECLIP’s Hit@1 reached 0.37, significantly outperforming BLEEP (0.26), CLIP (0.21), HisToGene (0.17), and ST-Net (0.16). Similarly, on spatialLIBD\_2\_HVG, HECLIP achieved a remarkable Hit@1 of 0.60, surpassing BLEEP (0.38), CLIP (0.37), HisToGene (0.33), and ST-Net (0.36). Moreover, as the $T$ value increased, the Hit@T values improve for HECLIP. Taken together, these benchmarking results demonstrate the robustness and superior predictive capability of HECLIP compared to alternative approaches.

\begin{table*}[]
\setlength{\tabcolsep}{0.05mm}{
\begin{tabular}{lllllllllllll} \hline \hline
\textbf{} & 
  \multicolumn{3}{c}{\textbf{GSE240429\_HVG}} &
  \multicolumn{3}{c}{\textbf{GSE240429\_HEG}} &
  \multicolumn{3}{c}{\textbf{GSE245620\_HVG}} &
  \multicolumn{3}{c}{\textbf{GSE245620\_HEG}} \\
\textbf{} &
  \textbf{Hit@3} &
  \textbf{Hit@2} &
  \textbf{Hit@1} &
  \textbf{Hit@3} &
  \textbf{Hit@2} &
  \textbf{Hit@1} &
  \textbf{Hit@3} &
  \textbf{Hit@2} &
  \textbf{Hit@1} &
  \textbf{Hit@3} &
  \textbf{Hit@2} &
  \textbf{Hit@1} \\ \hline
\textbf{HECLIP} &
  \textbf{0.9873} &
  \textbf{0.9627} &
  \textbf{0.8208} &
  \textbf{1} &
  \textbf{1} &
  \textbf{0.9978} &
  \textbf{0.9942} &
  \textbf{0.9772} &
  \textbf{0.8634} &
  \textbf{0.9998} &
  \textbf{0.999} &
  \textbf{0.9926} \\
\textbf{BLEEP} &
  0.9838 &
  0.957 &
  0.796 &
  \textbf{1} &
  \textbf{1} &
  \textbf{0.9978} &
  0.9914 &
  0.9637 &
  0.8253 &
  \textbf{0.9998} &
  \textbf{0.999} &
  \textbf{0.9926} \\
\textbf{CLIP} &
  0.9527 &
  0.9113 &
  0.7667 &
  \textbf{1} &
  0.9923 &
  0.9516 &
  0.917 &
  0.9003 &
  0.7551 &
  0.9536 &
  0.9433 &
  0.9177 \\
\textbf{HisToGene} &
  0.9399 &
  0.8669 &
  0.6723 &
  0.9637 &
  0.9191 &
  0.8367 &
  0.8679 &
  0.8458 &
  0.4135 &
  0.8777 &
  0.859 &
  0.5327 \\
\textbf{ST-Net} &
  0.9188 &
  0.8657 &
  0.6568 &
  0.9618 &
  0.9208 &
  0.8468 &
  0.8703 &
  0.8431 &
  0.4469 &
  0.8712 &
  0.8466 &
  0.5612 \\ \hline \hline
\textbf{} &
  \multicolumn{3}{c}{\textbf{spatialLIBD\_1\_HVG}} &
  \multicolumn{3}{c}{\textbf{spatialLIBD\_1\_HEG}} &
  \multicolumn{3}{c}{\textbf{spatialLIBD\_2\_HVG}} &
  \multicolumn{3}{c}{\textbf{spatialLIBD\_2\_HEG}} \\ 
\textbf{} &
  \textbf{Hit@3} &
  \textbf{Hit@2} &
  \textbf{Hit@1} &
  \textbf{Hit@3} &
  \textbf{Hit@2} &
  \textbf{Hit@1} &
  \textbf{Hit@3} &
  \textbf{Hit@2} &
  \textbf{Hit@1} &
  \textbf{Hit@3} &
  \textbf{Hit@2} &
  \textbf{Hit@1} \\ \hline
\textbf{HECLIP} &
  \textbf{0.8839} &
  \textbf{0.688} &
  \textbf{0.3688} &
  \textbf{0.9994} &
  \textbf{0.9793} &
  \textbf{0.5999} &
  \textbf{0.9987} &
  \textbf{0.976} &
  \textbf{0.5968} &
  \textbf{0.9993} &
  \textbf{0.9917} &
  \textbf{0.6214} \\
\textbf{BLEEP} &
  0.8321 &
  0.6636 &
  0.2585 &
  0.9987 &
  0.976 &
  0.5368 &
  0.9039 &
  0.7088 &
  0.3787 &
  0.9987 &
  0.976 &
  0.5968 \\
\textbf{CLIP} &
  0.8189 &
  0.6593 &
  0.2102 &
  0.9796 &
  0.964 &
  0.496 &
  0.8991 &
  0.6987 &
  0.3655 &
  0.9789 &
  0.9692 &
  0.5551 \\
\textbf{HisToGene} &
  0.8019 &
  0.7758 &
  0.1713 &
  0.8816 &
  0.8615 &
  0.4398 &
  0.873 &
  0.6456 &
  0.3333 &
  0.8439 &
  0.8009 &
  0.5178 \\
\textbf{ST-Net} &
  0.8113 &
  0.7963 &
  0.1628 &
  0.9013 &
  0.8688 &
  0.4513 &
  0.8695 &
  0.6428 &
  0.3618 &
  0.8415 &
  0.7998 &
  0.5231 \\ \hline \hline
\end{tabular}}
\caption{Comparative analysis of experimental results for $Hit@T$ metrics.}
\label{hk}
\end{table*}

\begin{figure*}[!htb]
\centering
\includegraphics[scale = 0.718]{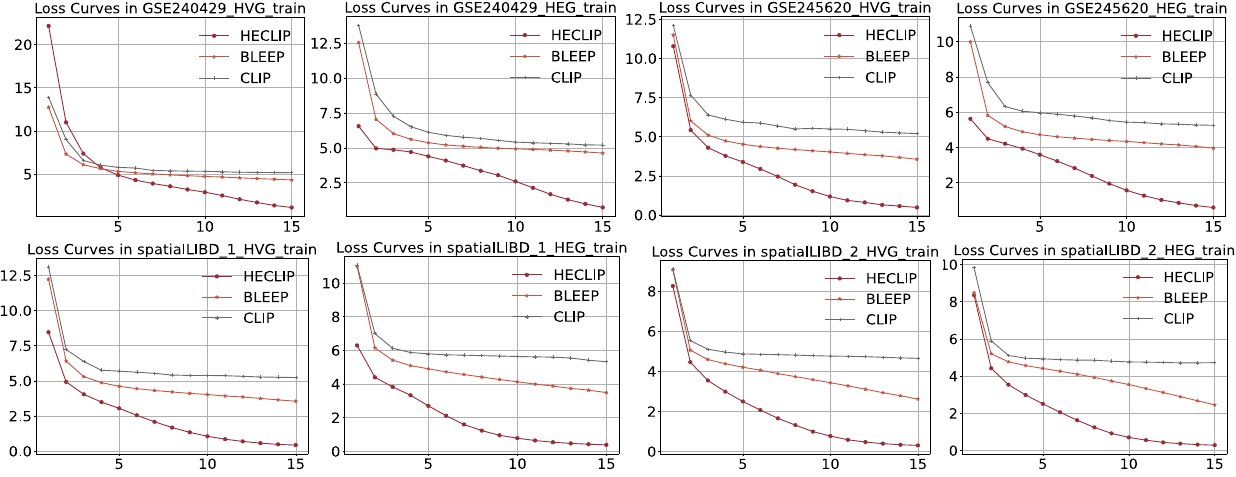}
\caption{Comparison of loss convergence across different methods in training stage.}
\label{loss}
\end{figure*}

\begin{figure*}[!htb]
\centering
\includegraphics[scale = 0.688]{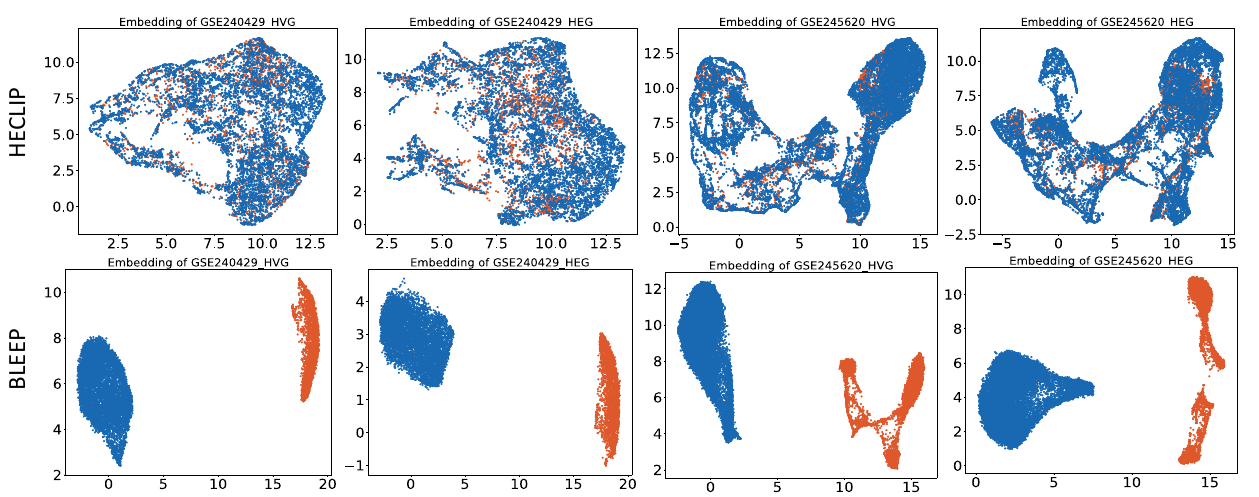}
\caption{UMAP of the bi-modality embeddings from HECLIP and BLEEP. The blue dots are the Reference set and the orange dots are the Query set.
}
\label{ebd}
\end{figure*}

\subsection{Loss convergence and embedding visualization}

Fig. \ref{loss} shows the reduction of loss over 15 epochs for different methods in training stage (testing stage is shown in Supplementary Fig. 2). HECLIP, optimized with an image-centric contrastive loss function, consistently outperforms both BLEEP and CLIP that rely on conventional loss functions. This advantage is evident in both training and test loss, highlighting the effectiveness of our tailored contrastive loss in enhancing model optimization and predictive capabilities.

Additionally, UMAP visualizations of the embeddings in GSE240429 and GSE245620 generated by HECLIP and BLEEP are presented in Fig. \ref{ebd} (spatialLIBD is shown in Supplementary Fig. 3), revealing significant differences in clustering patterns among the methods. The primary goal of these embeddings is to retrieve patches from the reference set that closely match those in the query set, which requires well-mixed and coherent representations. For BLEEP, the embeddings of the query and reference sets appear scattered, with limited integration between the two, indicating suboptimal alignment. In contrast, HECLIP’s embeddings exhibit a more cohesive and compact clustering, effectively mixing the reference and query sets. This demonstrates HECLIP’s ability to accurately capture similar patches. This cohesive embedding pattern is consistently observed across different datasets, underscoring the model’s robustness and reliability. These findings highlight the effectiveness of HECLIP’s unimodal contrastive loss in generating well-mixed, biologically meaningful embeddings, which significantly contribute to its superior overall performance.

\begin{table*}[]
\setlength{\tabcolsep}{1mm}{
\begin{tabular}{ccccc}  \\   \hline \hline  
\textbf{SSIM $\uparrow$}        & \textbf{Median}          & \textbf{Mean}         & \textbf{Median}         & \textbf{Mean}          \\ \hline 
                     & \multicolumn{2}{c}{\textbf{GSE240429\_HVG}}      & \multicolumn{2}{c}{\textbf{GSE240429\_HEG}}      \\
\textbf{HECLIP}      & \textbf{0.007049}        & \textbf{0.01889}      & \textbf{0.02382}        & \textbf{0.04594}       \\
\textbf{w/o loss} & 0.005791                 & 0.01753               & 0.02192                 & 0.04419                \\
\textbf{w/o data} & 0.005993                 & 0.01681               & 0.02241                 & 0.04444                \\ \hline 
                     & \multicolumn{2}{c}{\textbf{GSE245620\_HVG}}      & \multicolumn{2}{c}{\textbf{GSE245620\_HEG}}      \\
HECLIP               & \textbf{0.01105}         & \textbf{0.02737}      & \textbf{0.02537}        & \textbf{0.04421}       \\
\textbf{w/o loss} & 0.00996                  & 0.02641               & 0.02287                 & 0.04322                \\
\textbf{w/o data} & 0.01012                  & 0.02322               & 0.02391                 & 0.04326                \\ \hline 
                     & \multicolumn{2}{c}{\textbf{spatialLIBD\_1\_HVG}} & \multicolumn{2}{c}{\textbf{spatialLIBD\_1\_HEG}} \\
\textbf{HECLIP}      & \textbf{0.008819}        & \textbf{0.03224}      & \textbf{0.01877}        & \textbf{0.03452}       \\
\textbf{w/o loss} & 0.007891                 & 0.03178               & 0.01819                 & 0.03318                \\
\textbf{w/o data} & 0.007911                 & 0.03                  & 0.01796                 & 0.03268                \\ \hline 
                     & \multicolumn{2}{c}{\textbf{spatialLIBD\_2\_HVG}} & \multicolumn{2}{c}{\textbf{spatialLIBD\_2\_HEG}} \\
\textbf{HECLIP}      & \textbf{0.008607}        & \textbf{0.0309}       & \textbf{0.02854}        & \textbf{0.04782}       \\
\textbf{w/o loss} & 0.007928                 & 0.02858               & 0.02815                 & 0.04631                \\
\textbf{w/o data} & 0.007872                 & 0.02511               & 0.02801                 & 0.04474        \\   \hline \hline         
\end{tabular}}
\caption{Ablation experiment results of SSIM on all datasets. The upward arrow means the higher the better.}
\label{ssimab}
\end{table*}

\subsection{HECLIP accurately predicts biologically important genes}
HECLIP demonstrates high accuracy in predicting biologically significant genes, offering valuable insights into disease mechanisms and potential therapeutic targets. For example, HENMT1, an essential enzyme involved in piRNA methylation, is crucial for RNA stability and gene expression regulation. Disruptions in these mechanisms are implicated in various hepatic conditions, highlighting its importance in liver health \cite{eberhardt2007modulation}. Similarly, METTL11B plays a pivotal role in protein methylation, a key epigenetic modification that influences gene regulation and protein interactions. Aberrant methylation patterns mediated by METTL11B have been linked to liver disease progression, particularly hepatocellular carcinoma (HCC), where they contribute to oncogenesis and altered cellular signaling \cite{zhang2024mettl}.
PPIAL4A, a member of the cyclophilin family, is involved in protein folding, isomerization, and immune modulation. Cyclophilins are associated with liver inflammation and fibrosis, hallmark features of chronic liver diseases like cirrhosis and HCC. Cyclophilin inhibitors have shown promise in mitigating liver fibrosis and inflammation, underscoring their therapeutic potential \cite{naoumov2014cyclophilin}. Additionally, EIPR1 regulates endoplasmic reticulum (ER) stress, a process triggered by misfolded or unfolded protein accumulation. ER stress is a major contributor to liver diseases such as non-alcoholic fatty liver disease (NAFLD) and HCC \cite{chen2023endoplasmic}. 
These predictions exemplify HECLIP’s ability to bridge histological and transcriptomic data, enabling the discovery of genes with critical roles in disease mechanisms. By accurately identifying such biologically significant targets, HECLIP demonstrates its potential as a powerful tool for uncovering therapeutic targets and biomarkers in precision medicine.

\subsection{Ablation experiments of the HECLIP model}

In the ablation experiment, the effectiveness of HECLIP is thoroughly validated, as evidenced by the $RMSE$ and $SSIM$ results presented in Supplementary Table 1 and Table \ref{ssimab}. As shown in the tables, the HECLIP model consistently achieves relatively stable experimental outcomes in both \textbf{w/o loss} and \textbf{w/o data} settings.

\textbf{w/o loss} refers to the use of the original CLIP loss function. When using the original CLIP loss function, the model's performance on $SSIM$ and $RMSE$ is slightly worse compared to the HECLIP's image centric loss function. Specifically, for the GSE240429\_HVG dataset, the $RMSE$ median and mean values are 1.41 and 1.40, respectively, while the $SSIM$ median and mean values are 0.006 and 0.018. Similarly, in the spatialLIBD\_2\_HEG dataset, the $RMSE$ median and mean values are 1.36 and 1.35, and the $SSIM$ median and mean values are 0.028 and 0.046, respectively. While these metrics are lower than those achieved using the improved loss function, they remain superior to other methods such as ST-Net, highlighting the reliability of the improved loss function. This conclusion is further supported by the loss convergence in Fig. \ref{loss}.

\textbf{w/o data} indicates that no data augmentation strategy is employed. When data augmentation is not utilized, the model's performance on $RMSE$ and $SSIM$ is also worse. For instance, in the GSE240429\_HEG dataset, the $RMSE$ median and mean values are 1.3882 and 1.3656, and the $SSIM$ median and mean values are 0.02241 and 0.04444. Similarly, in the spatialLIBD\_2\_HVG dataset, the $RMSE$ median and mean values are 1.4201 and 1.4047, and the $SSIM$ median and mean values are 0.007872 and 0.02511.

Notably, employing both data augmentation techniques and the image centric loss function simultaneously yields the best performance. For example, in the GSE245620\_HVG dataset, the $RMSE$ median and mean values achieved by HECLIP are 1.3898 and 1.3743, while the $SSIM$ median and mean values are 0.01105 and 0.02737, respectively. These results underscore the synergistic benefits of integrating the image centric loss function with data augmentation, achieving superior predictive performance and stability across various datasets. We have also performed experiments of hyperparameter tuning, please refer to Supplementary Table 2 for the results.

\section{Conclusions}
This paper introduces HECLIP, an innovative CLIP-based model equipped with a specially designed unimodal contrastive loss function to enhance the representation capability of histological images. HECLIP exhibits strong scalability and adaptability, making it highly effective for contrastive learning tasks that align images with gene expression profiles. Comprehensive experiments across diverse datasets demonstrate that HECLIP consistently outperforms state-of-the-art models, delivering superior predictions with exceptional robustness and reliability.

\section{Competing interests}
No competing interest is declared.

\section{Author contributions statement}
Conceptualization: QW, BL, QS; Data collection: QW, QS; 
Software: QW; Formal analysis: QW, WC, QS; Writing— 
original draft: QW, WC, QS; Writing—review \& editing: 
QW, WC, BL, JS, GW, QS.

\section{Acknowledgments}
J.S. is supported by the National Library of Medicine of the National Institutes of Health (R01LM013771). J.S. is also supported by the National Institute on Alcohol Abuse and Alcoholism (R21AA031370 and U24AA026969), the National Institute of Health Office of the Director (OT2OD031919), the Indiana University Melvin and Bren Simon Comprehensive Cancer Center Support Grant from the National Cancer Institute (P30CA 082709), and the Indiana University Precision Health Initiative. Q.S. is supported by the National Institute of General Medical Sciences of the National Institutes of Health (R35GM151089). G.W. is supported by the National Institute of General Medical Sciences of the National Institutes of Health (1R35GM150460). This work partially used Jetstream2 \cite{hancock2021jetstream2} through allocation CIS230237 from the Advanced Cyberinfrastructure Coordination Ecosystem: Services \& Support (ACCESS) \cite{boerner2023access} program, which is supported by National Science Foundation grants \#2138259,  \#2138286,  \#2138307,  \#2137603,  and
\#2138296.

\bibliography{reference}

\end{document}